\documentclass[pre,a4paper]{revtex4}
\usepackage{graphicx} 
\usepackage{subfigure}
\usepackage{latexsym}

\def\sgn{{\rm sgn}}

\begin{document}

\author{M. Cosentino Lagomarsino} 
\affiliation{ FOM Institute for
  Atomic and Molecular Physics (AMOLF), Kruislaan 407, 1098 SJ
  Amsterdam, The Netherlands; Tel. +31 - (0)20 - 6081275 ; fax +31 -
  (0)20 - 6684106} 
\email[e-mail address ]{ cosentino-lagomarsino@amolf.nl }
\author{P. Jona} 
\affiliation{Politecnico di Milano, Dip. Fisica, Pza Leonardo Da Vinci
  32, 20100 Milano, Italy; Tel. +39 - (0)2 - 23996133 ; fax +39 - (0)2
  - 23996126} 
\email[ e-mail address ]{ patrizia.jona@polimi.it}
\author{B.  Bassetti}
\affiliation{{Universit\`a degli
  Studi di Milano, Dip.  Fisica, Via Celoria 16, 20100 Milano, Italy;
  Tel. +39 - (0)2 - 50317477 ; fax +39 - (0)2 - 50317480 
  and I.N.F.N., Via Celoria 16, 20133 Milano, Italy} 
\email[e-mail address ]{ bassetti@mi.infn.it } }

\title{Metachronal wave and hydrodynamic interaction for deterministic
  switching rowers}


\begin{abstract}
  We employ a model system, called {\it rowers}, as a generic physical
  framework to define the problem of the coordinated motion of cilia
  (the metachronal wave) as a far from equilibrium process.  Rowers
  are active (two-state) oscillators interacting solely through forces
  of hydrodynamic origin. In this work, we consider the case of fully
  deterministic dynamics, find analytical solutions of the equation of
  motion in the long wavelength (continuum) limit, and investigate
  numerically the short wavelength limit.  
  We prove the existence of metachronal waves below a characteristic
  wavelength. Such waves are unstable and become stable only if the
  sign of the coupling is reversed. We also find that with normal
  hydrodynamic interaction the metachronal pattern has the form of
  stable trains of traveling wave packets sustained by the onset of
  anti-coordinated beating of consecutive rowers.
\end{abstract}

\maketitle
\newpage

\section{Introduction}\label{sec:intro}

Cilia are hair-like extroflections of the cell membrane found in a
variety of species from protists to humans, which contain active
elements (molecular motors, filaments) acting as an internal drive
\cite{book}. Because of their size and typical velocities, the motion
of cilia is in the low Reynolds number regime.  Ciliary motion can be
divided in two stages, called \emph{power} and \emph{recovery}
strokes. The difference between the two is that in the power stroke a
higher portion of the surface of the cilium pushes against the fluid
compared to the recovery stroke so that, as in the breast stroke of
human swimming, the two effective viscous drags are different, and the
filament is able to propel the fluid \cite{Sleigh,purcell,BLA82}.
Cilia normally appear in arrays, and show coordinated wave-like
motion, referred to as \emph{metachronal wave}. The behavior of the
metachronal wave is thought to be strictly linked to the hydrodynamic
interactions between cilia \cite{GLL97,gueron1,gheber,gheber2}. The
question of how these collective motions are generated, from the
interplay between the internal, active degrees of freedom and the
external interaction is still open. The scope of this work is to
investigate this problem, using a simple deterministic model
containing very few parameters, consistent with experimental
observations and previous more detailed modeling, which can be easily
simulated and solved analytically in the limit of large wavelengths.

Modeling of cilia \cite{GH55,Br72,HB78} normally requires including
the infinite degrees of freedom of an inextensible line-like object,
its bending elasticity, and its interaction with the fluid (slender
body hydrodynamics), plus an active force, which can be imposed based
on physical observations or treated starting from the ``microscopic''
internal active degrees of freedom \cite{Mu90,julicher}.  The model we
present here (section \ref{model}) is extremely simplified and
economic in degrees of freedom. It is intended to be treatable
analytically. The cilia are represented by point particles, two-state
active oscillators which we call \emph{rowers}. The active force is
inspired by the switch mechanism introduced by Gueron and others
\cite{gueron1}. Planar or linear arrays of interacting rowers are
considered. In a previous work we used the same model to study the
role of noise \cite{sto_row}, proving that if the switch mechanism of
single rowers is purely stochastic the hydrodynamic interaction
generates metachronal waves which are statistically frustrated by the
presence of random fluctuations, but can be stabilized by the presence
of a short ranged coupling of the internal states, for example of
chemical origin. An alternative scenario we proposed was that the
presence of a coupling between position and transition frequency of
the single rower would lead to wave like solutions.  In this work we
would like to pursue this second possibility, in the limiting case
where the dynamics for the switch is governed deterministically by the
configuration of the rower, as in the geometric switch of Gueron
\emph{et alii}.


After an introduction of the model (section \ref{model}), we devote
the main body of the paper (section \ref{metach}) to the onset of
metachronal coordination. The discussion is divided in two parts. In
the first we discuss an analytical solution of the continuum limit of
our model equation, which enables to look for the onset of wave-like
patterns with large wavelengths. In the second part we look at the
short wavelengths through numerical simulations.
As we will show, rowers with a deterministic configurational switch
interacting hydrodynamically self-organize in patterns in which
nearest neighbor particles beat in anti-phase, and propagate trains of
wave packets with typical wavelength of a few particles. Only with an
effectively attractive interaction do long-wavelength wave-like
solutions appear.

\section{The model: rowers and evolution of the internal state} \label{model}

In the model we adopt, the movement of a single cilium is reduced to
that of a low Reynolds number rower that maximizes the effective drag
in its active phase (power stroke) and minimizes it in the passive,
recovery stroke.  More precisely, a rower is described by two degrees
of freedom, of which the first, \(f\) is translational and continuous
and represents a displacement from a reference position. It can be
thought of as the displacement of the center of mass of the filament
from an equilibrium position. The second, \(\sigma = \pm 1\), is
discrete and labels the internal state of the object, active or
passive.  The rowing direction is fixed, while the orientation can in
general be left open, to approach the problem of symmetry breaking in
the generation of fluid flow~\cite{sto_row}. The two states carry
different effective drags \( \gamma_{\sigma} = 1 + \epsilon \sigma\)
(with \(\epsilon >0\)) to the fluid, corresponding to different
surface impacts of the rower (different shapes of the filament) in the
two phases, together with different potentials (free energy
landscapes) \( V(f,\sigma)\), that generically describe different
active or relaxation forces felt by the cilium.  This is the
implementation of the so-called scallop theorem~\cite{purcell} at this
crude level of description, and makes it possible for a rower to
generate a net flow in the fluid.  There is no interaction between
rowers other than the force propagated by the presence of the fluid.
This force is modeled by the Oseen tensor for low Reynolds number
flows, appropriate in the case of cilia~\cite{purcell}.  The array of
cilia is modeled as a linear or planar lattice of rowers labeled by the index
\(n\), and the  configurations are specified by \(f_n, \sigma_n\).

\begin{figure}[htbp]
  \centering
  \includegraphics*[scale = 0.3]{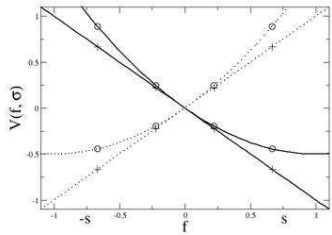}
  \caption{Potential \(V(f,\sigma)\) for \(k=0\) (+) and \(k =1\) (o).
    The dotted lines indicate \(\sigma = -1\), the solid lines
    \(\sigma = 1\). }
  \label{fig:potentials}
\end{figure}

This approach to the system contains a radical simplification in the
degrees of freedom of the object, a string with infinite degrees of
freedom, and of the active drive, generated by the collective behavior
of many molecular motors. This reduction enables us to carry an
analytical study. At our level of description, the substitution of
cilia with point particles does not change qualitatively the
interaction induced by the fluid.  However, it's a more delicate issue
to reduce the collective motion of molecular motors to a single
dynamic variable. We choose to maintain this variable discrete,
justified by the experimental observation that the motion of a cilium
is divided into two distinct phases and by previous, more detailed
modeling that suggested this picture~\cite{gueron1}
%
%
%

The evolution of a rower internal state can be modeled generically as
a stochastic process \cite{sto_row}, defined by the transition
frequencies between states. Here we analyze a limiting case where the
transition frequencies depend singularly on the configuration in a way
that reduces the dynamics to a deterministic one.  Essentially, a
rower contains a switch that alters instantaneously its internal state
when a particular limit configuration is reached.  The dynamics of the
internal switch is entirely local, in the sense that there is no
interaction of chemical origin between nearby rowers.  With these
choices, the evolution of \(\sigma_n\) may be represented by the
following equation containing a Dirac \(\delta\) distribution
\begin{equation}
  \label{eq:sigma}
   \frac{\partial \sigma_n(t)}{\partial t} = -2 \sgn(f_n(t)) 
   \left\vert\frac{df_{n}}{dt}\right\vert  \delta (f_n(t) -
   \textrm{s} \sigma_n(t)) 
\end{equation}
where \(\pm\)s are the switch points in correspondence of which the
discrete internal state is inverted. These parameters set the
amplitude of oscillation of a single particle and determine its window
of motion relatively to the driving potentials of the two states
(figure \ref{fig:potentials}).

Let's now turn to the evolution of the rower displacement and the
hydrodynamics.  Considering the fact that an overdamped motion follows
the maximum slope toward the minimum free energy and that we consider
no metastability, there are generically two possible qualitative
choices for the local conformation of the potentials in the two
internal states.  These can be linear, provided the system is far from
a minimum, or quadratic if it is close. Therefore, rescaling all the
constants that are not essential to our discussion (Stokes
coefficient, prefactors), we write:
\begin{displaymath}
  V(f,\sigma)= \frac{1}{2 k_{\sigma}}(k_{\sigma} f-\sigma)^{2} 
  -\frac{1}{2 k_{\sigma}}
\end{displaymath} 
where the parameter \(k_{\sigma} \in [0,1] \) determines the shape of
the potential (figure \ref{fig:potentials}).  For consistency
reasons, here s\(< 1\).

Thermodynamically,
\begin{itemize}
\item[-] if the switch is close to a minimum of the energy,
  \(k_{\sigma} = 1\) and we take a quadratic potential, the rower has
  time to dissipate completely its excess energy to the environment
  before it reaches the switch. The dynamics is a cyclical repetition of
  such relaxation processes
\item[-] if the switch is far from a minimum energy configuration,
  \(k_{\sigma} \to 0\) and the potential is linear. The switching
  process is faster than the thermalization time of the rower, that
  does not have time to dissipate all its energy. In this case at the
  switch the rower must undergo a collision-like process, which
  conserves the dissipation rate and the magnitude of the macroscopic
  velocity.   
\end{itemize}
The considerations above refer generically to the active mechanism of
model rowers, but leave aside the link with real cilia. If one wants
to give this drive a microscopic interpretation, it has to be in terms
of collective motions of the internal motors and elastic degrees of
freedom of the cilium. For example, in the linear potential scenario
one can imagine that motors attach/detach slowly generating a constant
force, while in the nonlinear one they attach simultaneously, giving
the cilium a well-defined minimum energy curvature, and they detach
collectively after reaching it. In what follows we will set
\(k_{\sigma} = k\) for simplicity.

The equation of motion for the rowers has to contain the hydrodynamic
interaction. We think of rowers as sources for the velocity field and
not as boundary conditions, which means introducing a (nondimensional)
coupling constant \(\alpha\) between the fluid and the rowers as a
substitute for the geometric constraints. \(\alpha\) is proportional
to the Reynolds number, or to the inverse of the kinematic viscosity.
To avoid complications, we do not take into account additional
boundary conditions, such as walls, but it is straightforward to
include them in the model. As the system is in a low Reynolds and
Strouhal number regime~\cite{landau}, it is possible to use the
regularized Oseen tensor~\cite{Doi} to eliminate the velocity field
and obtain an evolution equation for the sole rowers degrees of
freedom~\cite{sto_row}:

\begin{eqnarray}
  \frac {df_n(t)}{dt} & = &(1+\epsilon\sigma_n(t))\frac {\partial[ -
    V(f_n(t),\sigma_n(t))]}{\partial f_n} + \nonumber \\ 
  & + & \alpha \sum_{m \ne n}\Omega[{n},{m}] 
  \frac{\partial [- V(f_m(t),\sigma_m(t))]}{\partial
    f_m}
  \label{eq:1}
\end{eqnarray}

where \(\epsilon\) is the parameter that represents the difference
between the effective viscous drags in the two states and
\(\Omega[{n},{m}] = \frac{1}{r_{nm}} \hat{\mathbf{x}} \hat{\mathbf{x}}
(I + \mathbf{\hat{r}}_{nm} \mathbf{\hat{r}}_{nm})\) is the Oseen
tensor projected on the beating direction \( \hat{\mathbf{x}} \) of
the rowers. Strictly speaking, \( \Omega[{n},{m}]\) depends on \(f_n
,f_m\). However, to ease things in an analytical calculation, we
approximated it with a quantity that depends only on the relative
distance of the lattice sites, assuming that the oscillations are
small compared to this distance.  Most of our simulations, though,
were carried with the full \(\Omega\).

\section{Metachronal coordination} \label{metach}

The scope of this section is to establish whether hydrodynamic
interactions are sufficient to re-phase the rowers in absence of a
more direct coupling of chemical or mechanical origin.  
The oscillatory motion of a single rower in an array is guaranteed by
the structure of its equation of motion. A mean field description of
the array can be carried out considering the overall effect of the
velocity field generated on one rower by all the others. This
procedure is outlined in appendix \ref{app:meanf}, and leads to the
main result that a collection of rowers can generate a macroscopic
flow if \(\epsilon \ne 0\), and that, provided there is no intrinsic
orientation in the beating mechanism, symmetry will be broken.
However, in a description that goes beyond mean field, nothing can be
said about the beating time, which can in general vary with the
dynamics, so that the question can be restated as whether this
variability in the beating time is stabilized or disrupted by the
hydrodynamic interactions.  The problem is hard to approach
analytically due to the discontinuous nature of the switch \( \sigma\)
and the nonlinearities. However, the continuous limit of the model,
which describes the long wavelength behavior of the system, can be
approached analytically. The results obtained in this way can then be
compared with the numerical study of the discrete case.

From the point of view of solving the equation of
motion one has to investigate:
\begin{itemize}
\item[(1)] The existence of wave-like solutions. We define metachronal
  waves motions of the kind \(f_n(t) = f( t \pm \tau
  \mathbf{x_n})\), and simple metachronal waves those for which \(f\)
  (and thus \(\sigma\)) is a periodic function.
\item[(2)] The stability and attractivity properties of these solutions. 
\item[(3)] Their statistical weight in a macroscopic description of
  the system on large time scales. As one cannot establish \emph{a
    priori} its initial conditions, the metachronal solutions will be
  significant if the phase space volume of the initial conditions they
  attract is nonzero, and the relaxation time scales do not exceed a
  cutoff defined by the lifetime of the system.
\end{itemize}
In what follows we will be mainly concerned with the first two
points. The third point will be approached in general numerically
for systems of a few rowers (short wavelengths).  

\subsection{Metachronal pattern in the large wavelength limit. Continuous
  model} 

We will show that in the continuum limit of the equation of motion
\ref{eq:1} it is possible to find simple metachronal wave solutions
and study their stability analytically.  We can take the continuous
limit 
analyzing selectively metachronal solutions whose wavelength is large
compared to the spacing between rowers. Then \(f_n(t) = f(\mathbf{x_n},t)\)
becomes the continuous field \(f(\mathbf{x},t)\) and we can rewrite
the hydrodynamic interaction tensor \(\Omega[n,m]\) as
\((-\nabla^{2}+q^2)^{-1}\), where we incorporate also the possibility
of screening with inverse screening length \(q\).
With one inversion of this operator, the evolution equation \ref{eq:1}
for the continuous \(f\) can be written as
\begin{equation}
  (q^2 - \nabla^{2}) \left[\frac{\partial f}{\partial t}+
(1+ \epsilon \sigma)(k f - \sigma)\right] = - \alpha (k f - \sigma )
\label{eq:cont}
\end{equation}
The laplacian in the above expression is one dimensional, along the
fixed direction of beating \(\hat{\mathbf{x}}\) . We look for planar
metachronal wave solutions with the ansatz \( f = f( t -\tau x)\), so
that we can reduce to a 1+1 dimensional problem.  The transverse
hydrodynamic interactions are irrelevant as the rowers are constrained
to beat in one dimension, and the anisotropic terms can be absorbed in
the prefactor of the interaction tensor.  Calling \(y = t -\tau x\),
we can assume without loss of generality that \(y = 0\) is the
coordinate of a wave-front, where the switch \(\sigma\) has a jump.
This translates in the condition
\begin{displaymath}
   \sigma = \theta(-y)- \theta(y)
\end{displaymath}
where \(\theta\) is the Heaviside step function. The local
displacement \(f\) can be decomposed in the sum of power stroke and
recovery stroke parts, \( f_{+} \theta(-y) + f_{-}\theta(y)\), this
implies that
\begin{displaymath}
  \sigma f =f_{+} \theta(-y) - f_{-}\theta(y)
\end{displaymath}
With this procedure one obtains two linear third order differential
equations for \(f_-\) and \(f_+\), together with four joining
conditions for \(f_{\pm}\) and their derivatives in correspondence
with the jumps of \(\sigma\).  
Metachronal solutions can be constructed starting from the initial
condition \(y^{(0)} = 0\) and generating a succession of wave-front
coordinates \(y^{(1)}, y^{(2)}\ldots y^{(n)}\) imposing the joining
conditions above on the solutions of the third order differential
equations for \(f_{\pm}\). The iteration of this process, starting
from the initial conditions for \(f_{\pm}\) and its first and second
derivatives, or equivalently on the vector of the (two at the most)
independent arbitrary constants \((A_{\pm},B_{\pm})\) of the solution
of the differential equations, generates a flux in phase space,
described by an affine transformation.  The existence of a fixed point
of the succession \((A^{(n)}_{\pm}, B^{(n)}_{\pm})\) and its
attractive properties determine the nature and stability of the
constructed metachronal wave.  At every iteration, the \(y^{(n)}\)
must satisfy the relation \(f(y^{(n)}) = \pm \textrm{s} \).  Note
that despite every step involves linear operations, a strong
nonlinearity is introduced by the inversion for \(y^{(n)}\) of the
solutions of the differential equations.  It is also important to
stress that the solutions constructed in this way are in general not
periodic, and may have a domain of existence which is bounded in \(y\),
as after a number of iterations it could be impossible to invert for
\(y^{(n)}\).  More details on these calculations for the exemplifying
case \( k=1, \epsilon=0 \) are reported in appendix \ref{app:analytical}.

\begin{figure}[htbp]
  \centering
  \includegraphics*[scale=.3]{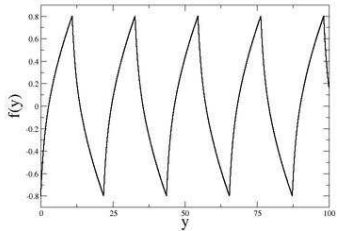}
  \caption{Analytical solution of the continuum equation
    (\ref{eq:cont}) computed for the case s \( = 0.8\), \(p = 0.01\), \(k =
    1\). The fixed point values of the parameters in this case are \(A_{\pm}
    = -0.435, B_{\pm} = -1.456\). }
  \label{fig:metach_an}
\end{figure}

\begin{figure}[htbp]
  \centering
  \includegraphics*[scale=.4]{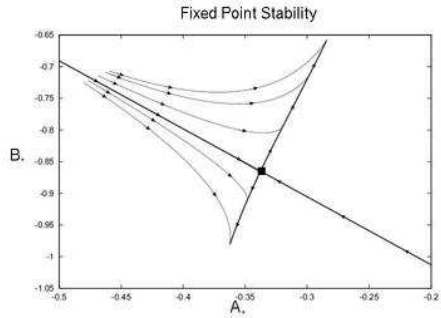} 
  \caption{Stability analysis for s\( = 0.3\), \(p = 0.05\), \(k =
    1\). The fixed point values of the parameters are
    \(A_{\pm} = \pm 0.337, B_{\pm} = \pm 0.865\). The fixed point is
    marked with a black square, and the lines with arrows indicate the
    direction of evolution of different initial conditions, showing
    that the fixed point is marginally stable. The trajectories are
    obtained iterating the transformation on \((A_{\pm},B_{\pm})\)
    starting from different initial conditions. The lines reported are
    Bezier interpolations of these (discrete) transformations.}
  \label{fig:sella}
\end{figure}

Our results can be summarized as follows. We find that, from the point
of view of the existence and stability of metachronal solutions,
\(\epsilon\) plays no qualitative role, so that the problems of
flow-generation and syncronization can be separated.  For this reason,
the discussion is independent on \(\epsilon\) and can be simplified
restricting to the case \(\epsilon = 0\).  A necessary condition to
find \( y^{(n)}\) is that s \( <1\), which gives a good consistency
test.
\begin{itemize}
\item[-] For quadratic potentials (\(k>0\)) the qualitatively relevant
  parameter is \( p = \frac{q+\alpha}{\tau^2}\). Fixing s, there is
  a critical value \(p_c\) such that for \(p>p_c\) there exist no
  solution. This set an upper critical wavelength \(\lambda_c\) for
  the metachronal waves. For \(0<p<p_c\), a fixed point exists,
  and the stability analysis gives a marginally stable saddle point in
  the planes of coefficients (\(A_{\pm},B_{\pm}\)). In other words
  there is one line (a region of zero measure) of stability in this
  plane, with runaway hyperbolic trajectories around it. We report
  this case in figure \ref{fig:sella} as an example. Finally, for
  \(p < 0\) the wave like solutions are always stable. However, this
  last condition implies \(p< 0\), therefore an effective
  ``attractivity'' of the hydrodynamic interactions. We will discuss
  below two cases that could lead to this situation.
\item[-] In the case \(k=0\) (linear potentials) the solution always exists.
  Stability analysis gives solutions that are always unstable (two
  eigenvalues \(>1\)) for \(p > 0\) and stable for \(p<0\).
\end{itemize}

In both cases we can conclude that metachronal solutions at long
wavelengths are unstable unless \(p < 0\), and the interaction becomes
attractive.  We would like to spend a few words on the possible
physical meaning of this change in sign, with an effectively
attractive interaction.  One first consideration is that interaction
between colloidal objects can be more complicated than how we
represent it, in presence of hydrodynamic effects or charge.  For
example, it has been speculated that the presence of a wall in
combination with a surface charge, all neglected in our model, lead to
effectively attractive potentials between colloids of like charges
\cite{brenner}.  Lubrication forces are another possibility, provided
that cilia are able to get sufficiently close to each other. For
rowers, this last condition is contrary to the assumption of small
oscillations in the theoretical calculation but can be tested through
simulations with the full Oseen tensor.
Coming back to the more orthodox case \(p > 0\), the results
above show that in the continuous limit simple metachronal waves (with
nearest neighbor rowers in phase) exist for all wavelengths (below a
characteristic one, for \(k\neq 0\)), but they are (marginally for
\(k\neq 0\)) unstable. These results are not sufficient to establish
in general the statistical behavior of metachronal waves, as the
hypotheses adopted here reduce the analysis to a particular class of
solutions and the local region of the phase space that surrounds them.
For negative \(p\) only we can conclude that wave-like solutions
attract all the initial conditions.  In the other cases, where the
solutions are not attractive, we have to look for other basins of
attractions in the phase space. One possibility is that these lie in
the shorter wavelength region, which is overlooked by the continuous
model. This motivates the numerical analysis of the next sections.

\subsection{Numerical simulations with many rowers}

\begin{figure}[htbp]
  \centering
  \subfigure[]{\includegraphics*[scale=.22]{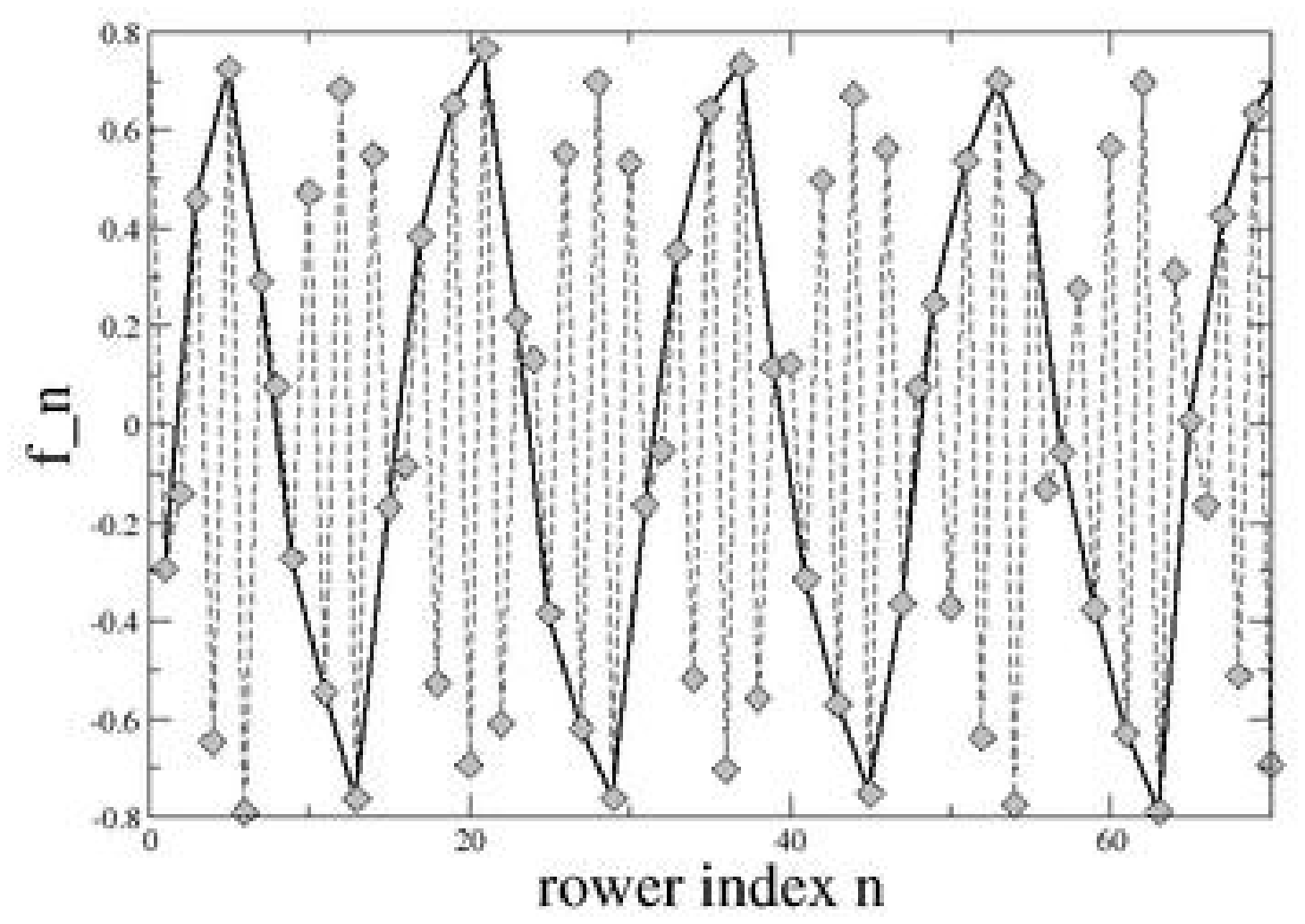}}
  \subfigure[]{\includegraphics*[scale=.22]{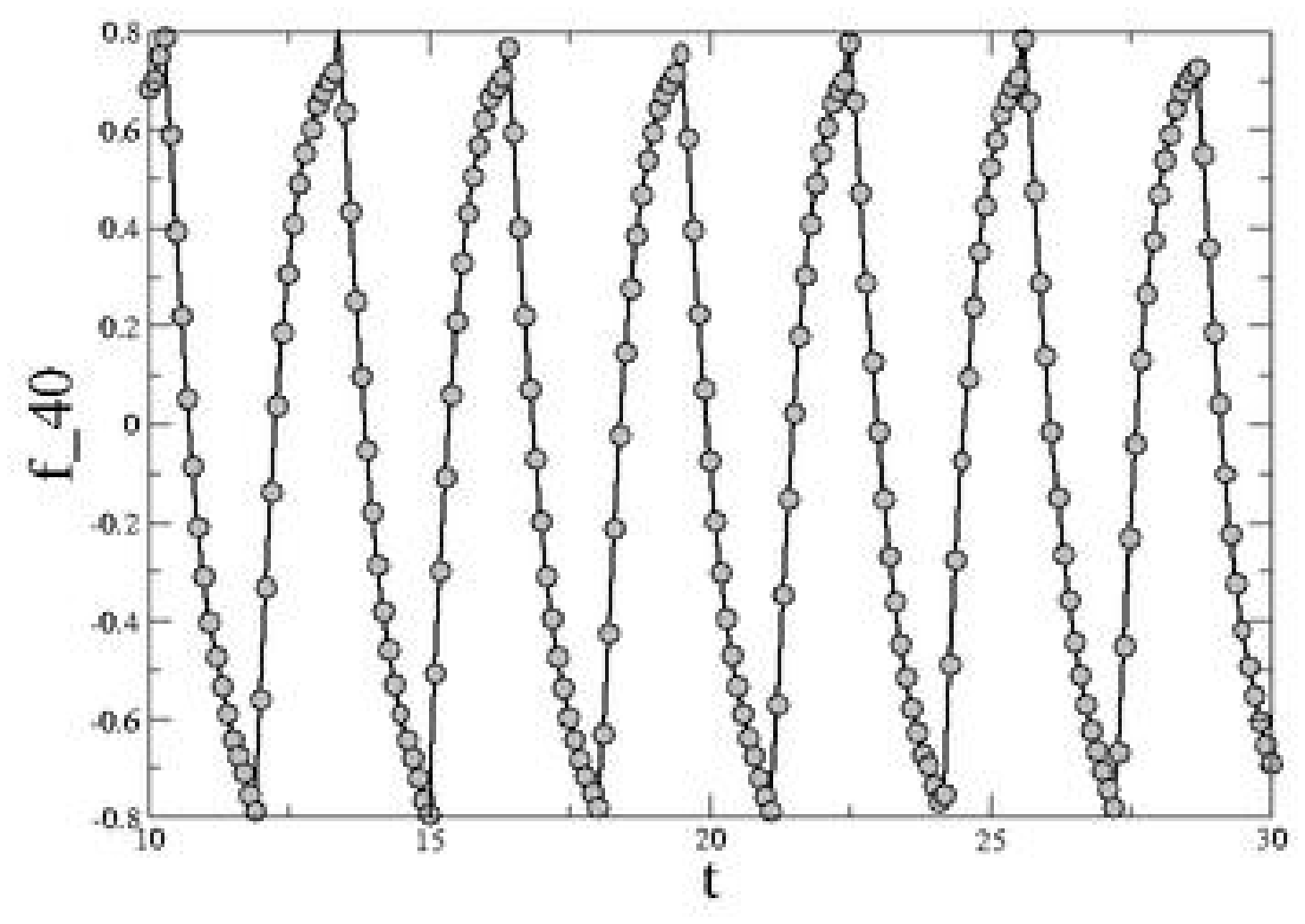}}
  \subfigure[]{\includegraphics*[scale=.22]{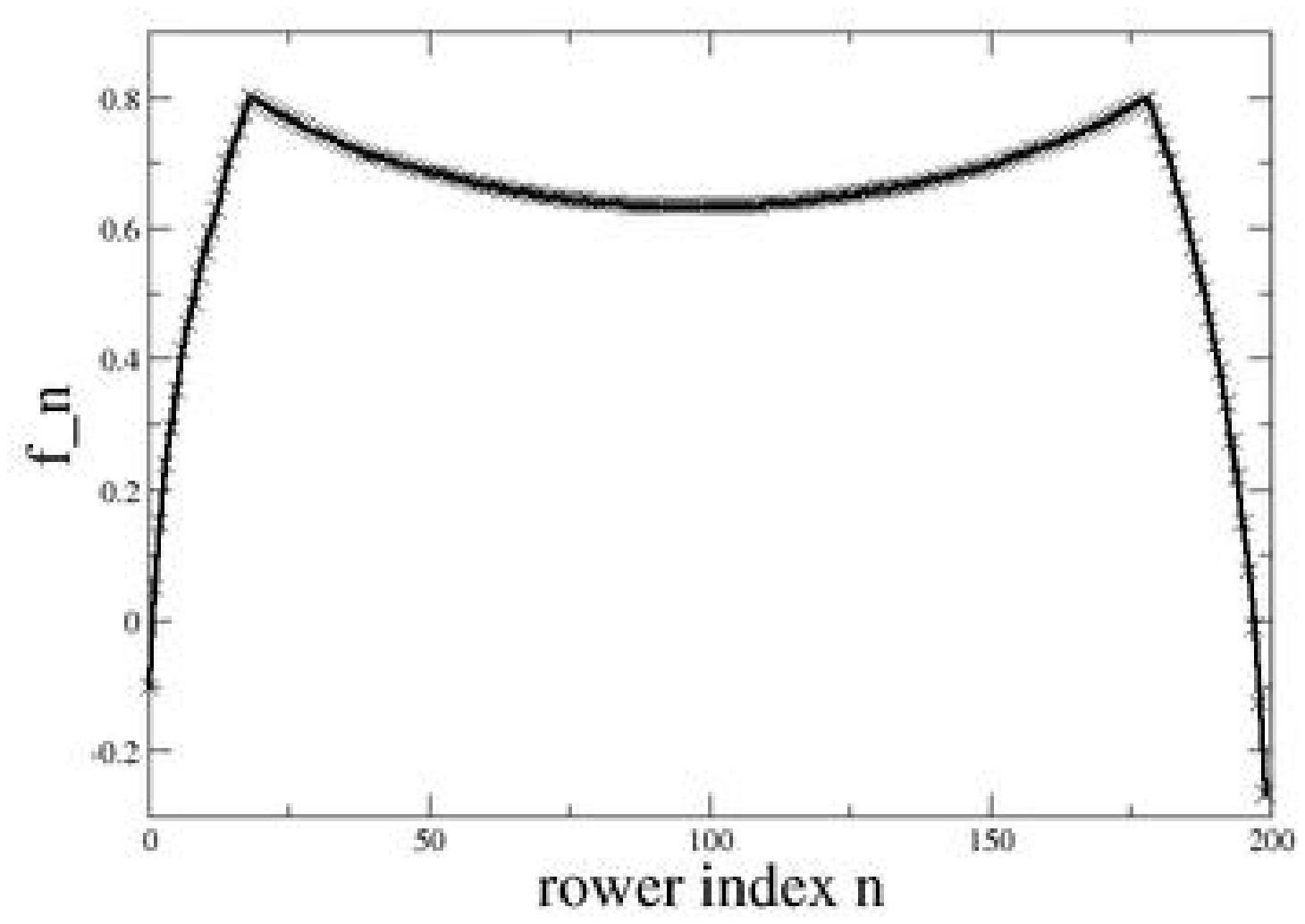}}
  \caption{ Metachronal solutions from numerical simulations of linear
    arrays of rowers interacting with the full Oseen tensor. The
    distance between successive rowers is 2 in our rescaled units and
    \(\epsilon = 0.25 \). (a) and (b): case \( p > 0\). \(\alpha =
    0.4, k = 1, 150\) rowers. In (a) a part the configuration \(f_n\)
    as a function of the rower index is shown. The dashed line shows
    the actual configuration, while the solid one, obtained connecting
    the points relative to odd rowers, outlines the shape of the wave
    packet. (b) represents the time evolution \(f(t)\) of the 40th
    rower of this array. (c): case \( p < 0\). \( \alpha = -0.1, k =
    1, 200\) rowers. The configuration \(f_n\) shows a long wavelength
    metachronal wave. The part of the configuration for \(i > 100\) is
    mirror symmetric, due to the same pattern traveling in the
    opposite direction}
  \label{fig:nummany}
\end{figure}

A confirmation of these results, and more insight on the behavior of
the system comes from numerical simulations of linear and two
dimensional arrays of many (50-250) rowers.  These were run starting
from random initial conditions, with nearest neighbor Oseen-like
interactions, the simplified interaction tensor \(\Omega [n ,m] \) and
the full \(\Omega [f_n ,f_m ] \), robustly showing the same
phenomenology in all cases. In particular, for linear arrays with \(
k>0\):
\begin{itemize}
\item[-] For \( p>0\) the relaxed solutions look like trains of
  traveling wave packets where nearby rowers show anti-coordinated
  motion.  In the patterns observed in the simulations the packets
  have typically a characteristic length of the order of 10 particles.
  Their traveling direction coincides almost always with the beating
  direction of single rowers.
\item[-] For \( p<0\) long wavelength traveling solutions appear.
  These always have a wavelength that is exactly the size of the
  array.  For every solution of this kind traveling in one direction,
  there is another one in the opposite one, leading to mirror symmetric
  standing waves. The same solutions are found if \(\alpha\) is kept
  positive, and a nearest neighbor attractive interaction is added.
\end{itemize}
This is exemplified in figure \ref{fig:nummany}. For linear arrays
with \(k = 0\) wave-like solutions appear only for \( p < 0\). Two
dimensional arrays show the same qualitative behavior. This makes
the wave patterns propagate in directions that are different from the
beating direction of the rowers (non-simplectic, in the
language of metachronal waves).

We can give a heuristic argument based on the switch mechanism to
account for this anti-phase coordination.  Let's consider three
consecutive rowers X,Y,Z, and suppose they are in phase. Immediately
after X reaches the +1 switch it feels a strong negative force due to
the change in the potential, which is propagated mainly to Y and much
less to Z.  Provided Y was going toward the same switch, it will be
slowed down, and therefore de-phased with respect to X. If we imagine
now that X,Y,Z are in anti-phase and A reaches the +1 switch. Now the
force felt by Y (and much less by Z) will be driving it toward the -1
switch, reinforcing the dephasing. As a result, the anti-phase beating
between X and Y is more stable than the coordinated motion. According
to this argument, this will be the case in all the instances where the
influence of nearest neighbors is stronger than that of far away ones.
More formally, one can find a reason for this behavior in the
analytical solution of the continuum equation. Configurations where
rowers, interacting with the normal Oseen tensor have opposite phase
are overlooked by the continuum limit.  However, roughly speaking,
these configurations lead effectively to a change of sign of
\(\alpha\) in the continuum limit equation, as one can easily see
imposing \(f_{\textrm{\tiny even}} = - f_{\textrm{ \tiny odd}} \) in
 equation \ref{eq:1}.  Therefore one can think that the
stable waves found for \( p<0\) in the continuum model have a trace of
this anti-phase behavior.



\subsection{Stability of the metachronal wave at short wavelengths.} 

\begin{figure}[htbp]
  \centering
  \subfigure[]{\includegraphics*[scale=.3]{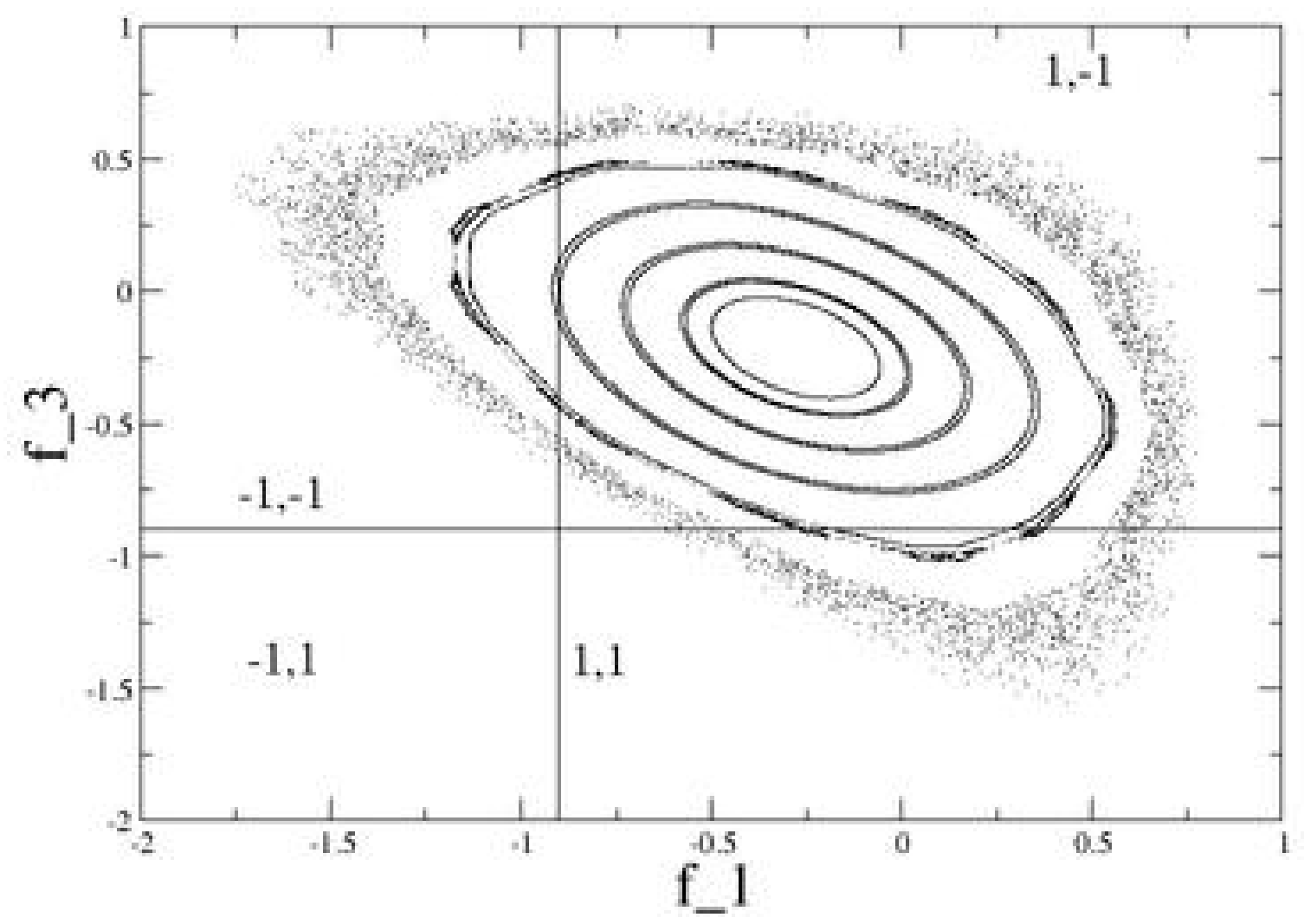}}
  \subfigure[]{\includegraphics*[scale=.3]{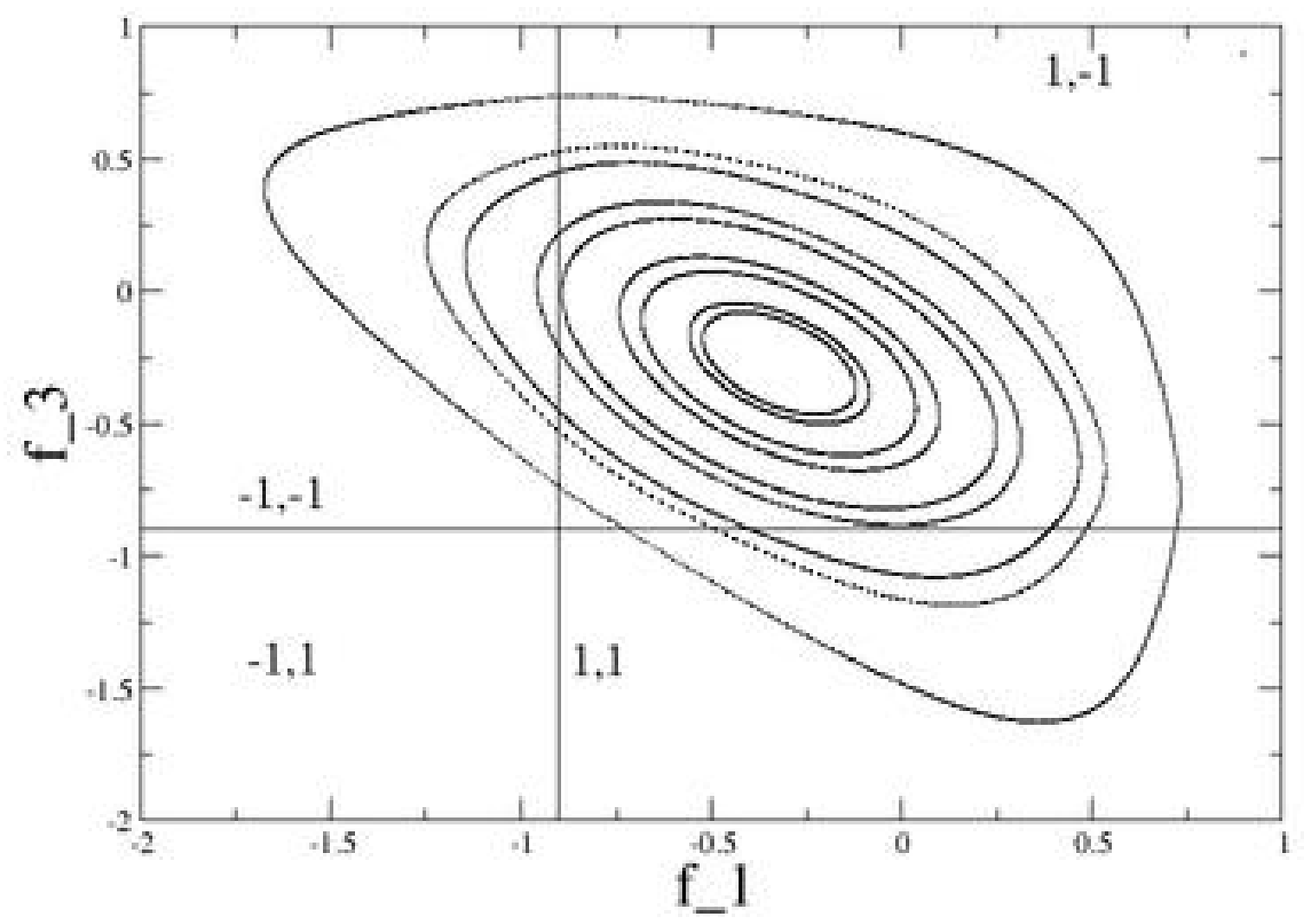}}
  \caption{Poincar\'e map for the case \(k = 0\), three rowers,
    periodic boundary conditions. Here, \(\epsilon = 0.02\) and s\(=
    0.9\). The allowed configurations are confined in the square
    region [-s,s]\(\times\)[-s,s]. However, to increase clarity, the
    figures were divided in four regions, corresponding to the
    different values of the two switches, which were ``folded
    outside'', so that the axes correspond to the actual configuration
    of the rowers after a mirror reflection dependent on the state of
    the switches.  (a) \( \alpha = 0.1\).  (b) \(\alpha = 0.01\).  }
  \label{fig:centro}
\end{figure}

\begin{figure}[ptb]
  \centering
  \subfigure[]{\includegraphics*[scale=.38]{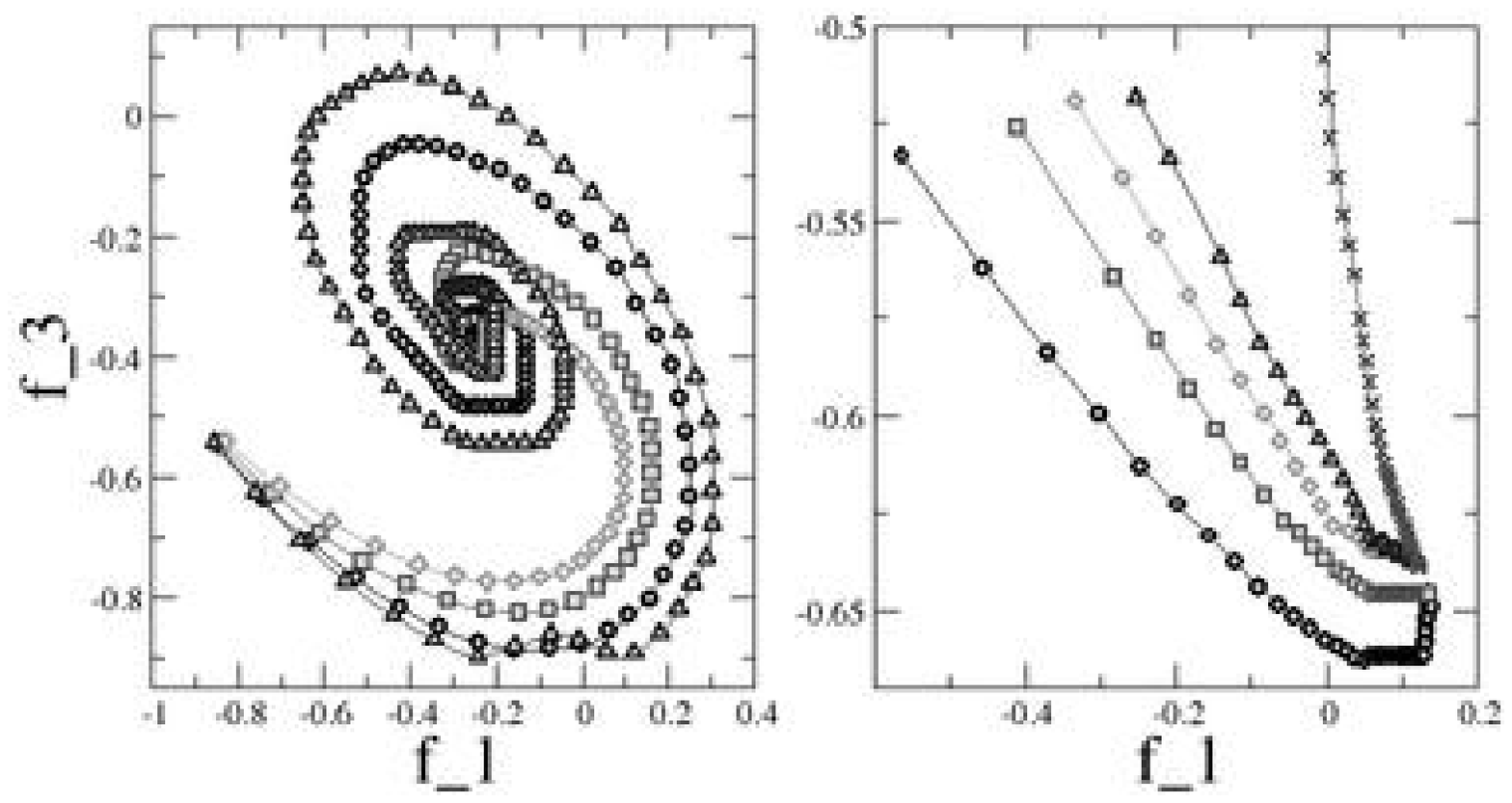}}
  \subfigure[]{\includegraphics*[scale=.22]{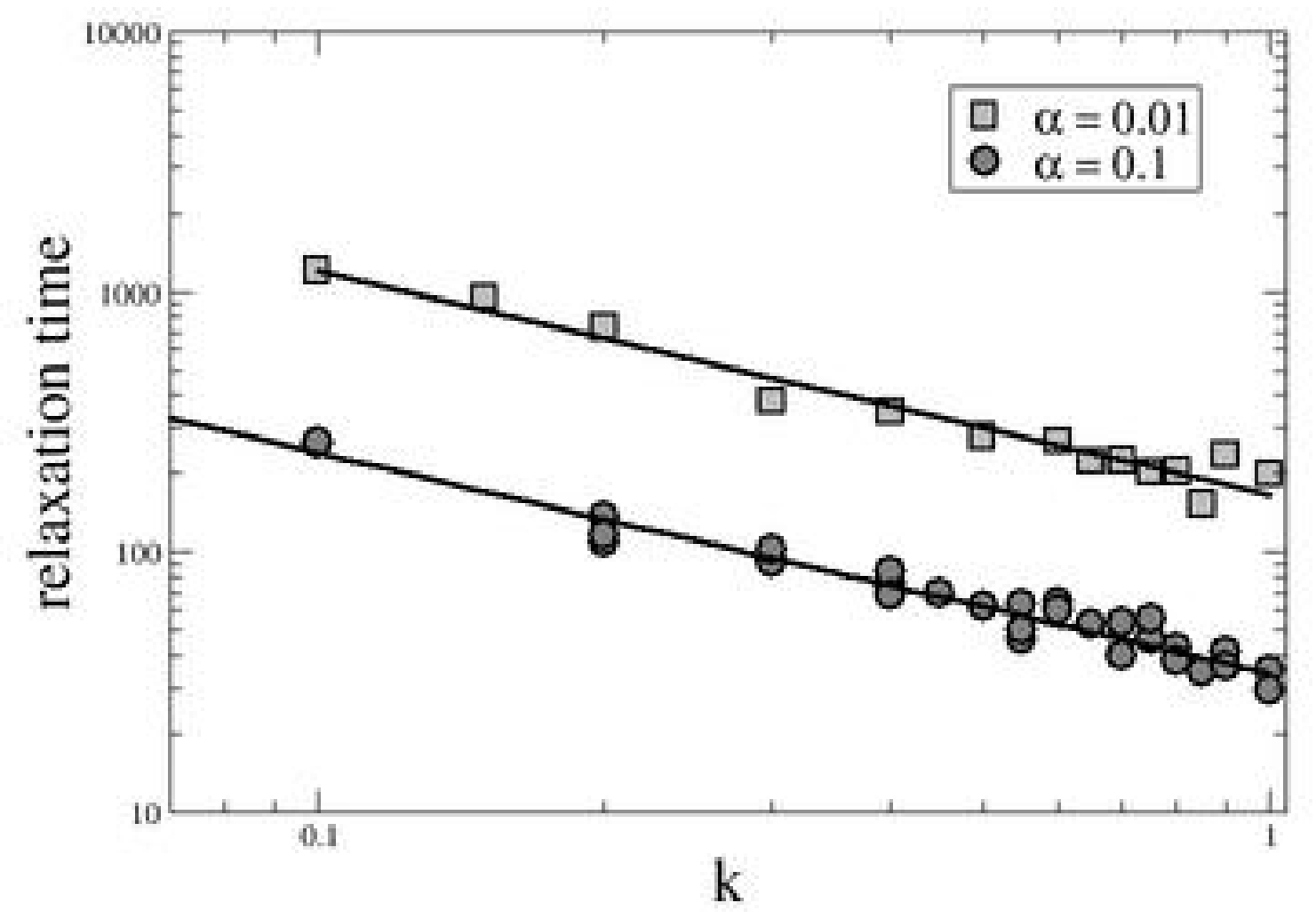}}
  \subfigure[]{\includegraphics*[scale=.22]{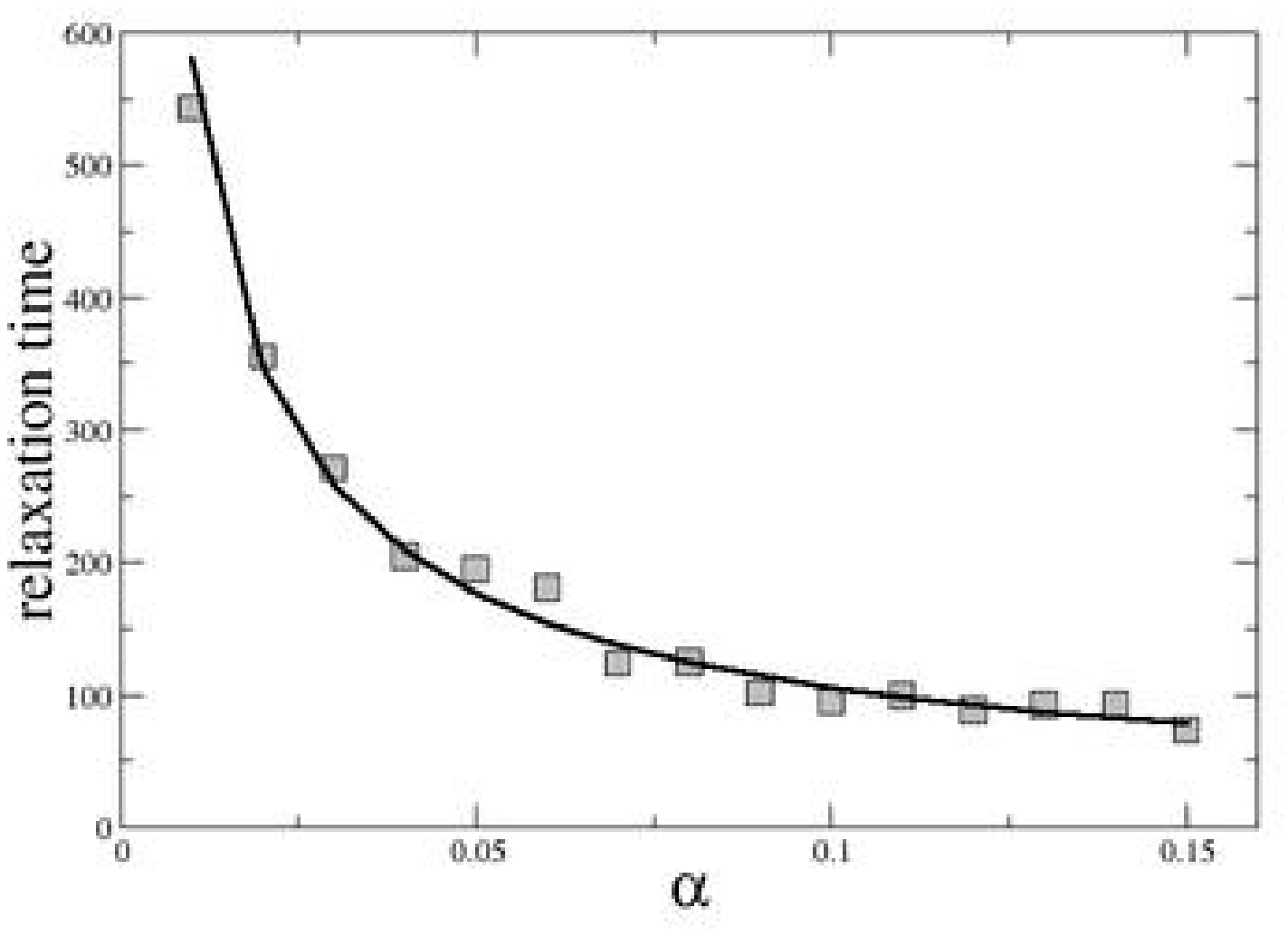}}  
  \caption{(a) Poincar\'e maps for \(k \ne 0 \), three rowers, \( \epsilon =
    0.02 \). Left: \(\alpha = 0.01\) attracting trajectories for
    different values of \(k\).  \(\triangle\) \(k =0.15\). O \(k =
    0.2\). \( \Box\) \(k = 0.3\). \(\Diamond\) \(k = 0.4\). Right:
    different initial conditions for \(k = 1\). (b) log-log plot of
    the relaxation time as a function of \(k\) for \( \alpha = 0.01
    (\Box) \) and \(\alpha = 0.1\)  (O). The power law fits (solid
    lines) yield an exponent of \(0.85\). (c) Relaxation time as a
    function of \(\alpha\) for \(k = 0.3\). The power law fit gives
    the exponent \(0.74\).}
  \label{fig:attract}
\end{figure}

To analyze the system at shorter wavelengths we ran numerical
simulations of linear arrays of a few rowers with periodic boundary
conditions, choosing to truncate the interaction to nearest neighbors.
We interpret the boundary condition as a constraint on the wavelength
of the resulting wave patterns. For up to four rowers, we analyzed the
system using Poincar\'e maps.  This study leads to more general
considerations on the statistical weight of metachronal solutions. The
maps are obtained graphing the positions of two rowers when a reference
one reaches the switch over many cycles of the motion and for different
initial conditions. If the resulting graph is, or converges to a
single point, the motion is coordinated. If it is a closed orbit the
coordination is quasi-periodic, which means that the motion in general
does not look like a traveling wave. Finally, if the resulting graph
is a random scatter plot, the motion is chaotic. Indeed, the system
can be either chaotic or quasi-periodic if the potentials are linear,
(\(k = 0\)), as can be seen in figure \ref{fig:centro}. The phase
space volume of the quasi-periodic region delimited by the concentric
loops is therefore a measure of the statistical weight of
quasi-periodic solutions. The volume of this region decreases with
increasing \(\alpha\).
The case \(k \ne 0\) is somewhat different. Here, there is no chaos,
and the stable tori become attractors to a fixed point of the
parameter space (figure \ref{fig:attract}, (a)). This means in
principle that all the initial conditions lead to pattern formation.
Therefore, the relevant parameter becomes the relaxation time. This
diverges as \( k\rightarrow 0\) as a power-law with exponent \( b
\simeq 0.85\) and as \( \alpha \rightarrow 0 \) with exponent \( c
\simeq 0.74 \). 
The curvature of the potential, which as we discussed has to do with
the activity of the microscopic active degrees of freedom, seems
therefore to be important in determining the organization properties
of the system. On the other hand, the phenomenology observed for
different values of \(\alpha\) is qualitatively consistent with what
observed experimentally for arrays of cilia beating in fluids with
varying viscosity~\cite{gheber2,gheber}.

\section{Overview and conclusions}

We have presented a simple model system of two-state low Reynolds
number oscillators called rowers as a generic framework for the
problem of cooperation of cilia. The dynamics adopted in this work,
specified setting the transition rates between the two potentials, is
entirely deterministic, determined by a switch mechanism coupled to
the configuration.  
We solved analytically for wave like solutions the continuum, long
wavelength, limit of the equation of motion for an array of rowers
with hydrodynamic interaction and we analyzed the stability of the
solutions, confronting with results from numerical simulations.
Finally, we analyzed through Poincar\'e maps the phase space dynamics
of systems of a few rowers, to study their behavior at short
wavelengths.  

Our most important result is that metachronal patterns exist at all
wavelengths (below a characteristic one, for \(k\neq 0\)), but long
wavelength solutions are (marginally for \(k\neq 0\)) unstable.  The
stable patterns have the form of consecutive wave-packets where
nearest neighbor oscillators are in anti-phase, propagated with
constant speed, with a characteristic length of a few rowers. We
showed that the statistical weight of these solutions can be
determined numerically imposing an upper cutoff on the wavelength of
the pattern.  Only in the presence of a reversed coupling constant,
can long wavelength metachronal solutions be stable. We proposed two
possible physical reasons for this reversal in sign.
Deterministic switching rowers, as two state oscillators, show a rich
and unusual phenomenology, of which we could explore a number of
aspects. Their behavior is in many ways opposite to our usual notion
of oscillations, starting from the fact that no normal modes can be
defined, but the oscillators self-tune to a chosen frequency
determined by the characteristic relaxation times in the two states,
much as in systems close to a Hopf bifurcation \cite{julicher}. 

Comparing the behavior our abstract entities with that of real or
model cilia, the first puzzling question seems to lie in the anti-phase
motion. As discussed, a solution of this could lie in a short ranged
interaction with a different origin. One good candidate for this are
lubrication forces, as real cilia can be really close to each other.
Also, a short ranged synchronization between the switches of chemical
origin could lead to the same result, consistently with the scenario
proposed in our previous work. The relevant parameters in our
discussion are the stiffness of the potential \(k\) and the
hydrodynamic interaction coupling strength \(\alpha\). The first is
related to the internal active degrees of freedom, which are hard to
access experimentally, while the second can be used for a qualitative
comparison of our results with experiments where arrays of cilia are
observed beating in fluids with varying
viscosity~\cite{gheber2,gheber}.
One other question is the relation with more detailed models of cilia
and their internal drive, in particular with the geometric switch
model of Gueron and collaborators \cite{GLL97, gueron1}. Rowers, with
their few degrees of freedom constitute a system much more under
control than filaments to test. We can conclude that generically
simple hydrodynamic interaction does not synchronize but
anti-synchronizes nearest neighbor rowers, so that, if filamentous
objects are to be synchronized by a similar mechanism, an extra (to be
found) ingredient is needed.

\begin{acknowledgments}
The authors would like to thank Davide Rossi for his contribution to
this work. 
\end{acknowledgments}

\appendix

\section{Mean field approach, symmetry breaking} \label{app:meanf} 
\begin{figure}[htbp]
  \centering
  \includegraphics*[scale=0.3]{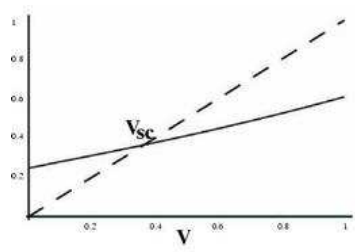}      
  \caption{Selfconsistent velocity calculation. The right hand side
    (solid line) and the left hand side (dashed line) of equation
    \ref{eq:self} are plotted as a function of \(v\) in the graph
    above. The intersection between the two lines yields
    \(v_{\textrm{sc}}\). The values of the parameters for the curves
    shown are \(k=0.8, \epsilon= 0.25, \textrm{s}= 0.8\), and \(\alpha
    \sum_{n \ne 0}\Omega_{0,n} = 0.2\). With these values,
    \(v_{\textrm{sc}} \simeq 0.365\) in nondimensional units. }
  \label{fig:sc}
\end{figure}

It is possible to estimate the magnitude of the characteristic
beating times and the macroscopic speed generated on the fluid without
having to solve explicitly the equations of motion, using simple
self-consistency considerations~\cite{sto_row}. This is done writing
the equations of motion (\ref{eq:sigma} and \ref{eq:1}) for a single
rower in a constant external effective velocity field \(\mathbf{v}\)
along its beating direction and supposing that this velocity is
generated by the effect of the surrounding rowers. Equation 
\ref{eq:1} looks then like
\begin{displaymath}
  \frac {df_0(t)}{dt}  = (1+\epsilon\sigma_0(t))\frac {\partial[ -
    V(f_0(t),\sigma_0(t))]}{\partial f_0} + v
\end{displaymath}
where we labeled conventionally with the index 0 the rower around
which we do the selfconsistent calculation.
It is then straightforward to calculate the beating times \(t_{\pm}\)
for this single rower, defined as the times required to go from -s to
+s and back respectively:
\begin{displaymath}
  \begin{array}{c}
    t_+ = [k (1+\epsilon)]^{-1} ln \left[\frac{(1+\epsilon)(1+k \, 
        \textrm{s})+v} {(1+\epsilon)(1-k \, \textrm{s})+v}\right] \\ 
    \\
    t_- = [k (1-\epsilon)]^{-1} ln \left[\frac{(1-\epsilon)(1-k \,  
        \textrm{s})-v} {(1-\epsilon)(1+k \, \textrm{s})-v}\right]   
  \end{array}
\end{displaymath}
These quantities can be used to determine self-consistently the
absolute value of the ``macroscopic'' fluid velocity \(v\), taking
into account the average force exerted by the single rower on the
fluid in one cycle.
\begin{displaymath}
  v_{\textrm{sc}} = \left[\alpha \sum_{n \ne 0}\Omega_{0,n}\right]
    \frac{1}{T}\int_{0}^{T} -\frac{\partial V(f,\sigma)}{\partial f}dt
\end{displaymath}
where the total period \(T = t_+ + t_-\) depends on \(v\). Therefore
\begin{equation}
  v_{\textrm{sc}} =  -  \frac{1}{T} 
  \left[\alpha \sum_{n \ne 0}\Omega_{0,n}\right] 
  \left[v_{\textrm{sc}} 
    \left(\frac  {t_+}{1+\epsilon}+\frac{t_-}{1-\epsilon}\right) + 4 \epsilon \,
  \textrm{s}\right]     
\label{eq:self}
\end{equation}
In this expression the summed hydrodynamic propagator simply plays the
role of a multiplicative constant, and s determines the
self-consistent value for the velocity (see figure \ref{fig:sc}). For
\(\epsilon=0\), \(v_{\textrm{sc}}=0\), and the scallop theorem is found again
in the description of a collection of rowers.
The same argument yielding a non zero self-consistent velocity can be
extended to the case of one - or a collection of- rowers whose rowing
direction is not fixed. This is obtained taking reflection symmetric
potentials \(V(f,\sigma) = V(-f,\sigma) \), one of which with a single
minimum for \(f= 0\), the other with a double well. A thermal noise
needs to be added for the problem to be well posed. Rowers are able to
break spontaneously the symmetry to generate a flow in the fluid
\cite{sto_row}. For real cilia, the problem of symmetry breaking can
be relevant in the context of generation of left-right asymmetry
through nodal flow in vertebrate embryos~\cite{nonaka}.


\section{Example of explicit solution of the continuous model and its
  stability} \label{app:analytical}

We will solve equation \ref{eq:cont} analytically with the ansatz \( f
= f(x - \tau t) \) on the solution. For simplicity we can restrict
ourselves to the case \(k = 1\), \( q = 0\), \( \epsilon = 0\), as the
general case carries no further conceptual complication.  Calling \( y
= x - \tau t\), the equation reads:
\begin{equation}
  \label{eq:y}
  -\tau^2 (f^{'''} + f^{''}) + \alpha (f - \sigma) + \tau^2
   \sigma^{''} = 0
\end{equation}
where \('\) indicates derivatives with respect to \(y\).
For a transition of \(\sigma\) from 1 to -1 at the wave-front \(y =
0\), the right joining conditions are, as already discussed,
\begin{eqnarray}
   \sigma = \theta(-y)- \theta(y)  \label{eq:joina} \\
   f=f_{+} \theta(-y) + f_{-}\theta(y)  \label{eq:joinb} \\
   \sigma f =f_{+} \theta(-y) - f_{-}\theta(y)  \label{eq:joinc}
\end{eqnarray}
where \(\theta\) is the Heaviside step function, and \( f_{\pm} =
f_{\pm}(y)\). Analogue expressions hold for the transition \( -1
\rightarrow 1\). The decompositions above generate two linear
ordinary differential equations for \(f_{\pm}\). 
\begin{equation}
  \label{eq:fpm}
  f_{\pm}^{'''} + f_{\pm}^{''} - p f_{\pm} = \mp p
\end{equation}
with \(p =  \frac{\alpha}{\tau^2}\). Moreover, the same conditions
\ref{eq:joina}, \ref{eq:joinb}, \ref{eq:joinc} and their derivatives can
be substituted in equation \ref{eq:y}, obtaining an expression
containing terms in \(\theta(\pm y)\), \(\delta(y)\) and its
derivatives.  Equating all the terms to zero one obtains three joining
conditions. That is,
\begin{eqnarray}
  (f_+ - f_-)\vert_{\textrm{switch}} = 0; \\ 
  (f^{'}_+ - f^{'}_-)\vert_{\textrm{switch}} = \frac{2}{3}; \\
  (f^{''}_+ - f^{''}_-)\vert_{\textrm{switch}} = -\frac{4}{9}
  \label{eq:join2}
\end{eqnarray}
Here, and more generally in the case \(\epsilon = 0\), the two
equations \ref{eq:fpm} are the same with the identification \( -f_- =
f_+\).  The solution of equation \ref{eq:fpm} is easily obtained;
\(f_{\pm} = \pm 1\) is always a particular solution, and one has to
solve the characteristic equation \( z^3 + z^2 = p \). This admits
three real solutions for \( 0 < p < \bar{p} \), and one for \(p < 0. p
> \bar{p}\), \(\bar{p}= 4/27\).  We will analyze in detail here the
case with three solutions (\(-z_1, -z_2, z_3\)), with \(z_1, z_2, z_3
> 0 \) . In this situation,
\begin{displaymath}
  f_{\pm}(y) = \pm \left( 1+  A_{\pm} e^{-z_1y} + B_{\pm} e^{-z_2y} +
  C_{\pm} e^{z_3 y} \right)
\end{displaymath}
The constant \(C_{\pm}\) can be eliminated using the condition \(
f_{\pm}(0) = \mp \textrm{s} \), meaning that after the first jump the
rower is located at the switch. The next step is to evolve this
solution up to a certain \(y_{\pm}\) where the next switching event
will take place, imposing that
\begin{equation}
 f_{\pm} (y_{\pm}) = \pm \textrm{s}
\label{eq:ypiu} 
\end{equation}
\(y_{\pm}\) is obtained inverting this last expression, and has to
satisfy the joining conditions \ref{eq:join2} for the next
``piece''. For example, supposing we start from state \(\sigma = 1\) 
\begin{displaymath}
  (f^{'}_+(y_+) - f^{'}_-(0)) =  \frac{2}{3} \ ; \ \ 
  (f^{''}_+(y_+) - f^{''}_-(0)) = - \frac{4}{9}
\end{displaymath}
This gives a linear transformation \( (A_+,B_+) \rightarrow
(A_-,B_-)\). 
A complete solution can be constructed iterating this procedure. This
solution is in general non-periodic, as \(y_{\pm}\) may vary at every
step, and also cease to exist.
The equations that determine \(y_{+}^{(n)}\) and \(y_{-}^{(n)}\) at the
\(n\)-th step can be written as 
\begin{equation}
    A_{\pm}^{(n)} H_{(1,3)} (y_{\pm}^{(n)}) + B_{\pm}^{(n)}
    H_{(2,3)}(y_{\pm}^{(n)}) -(1+\textrm{s}) E_{3}(y_{\pm}^{{(n)}})
    +(1-\textrm{s})=0   
\label{eq:t+-}
\end{equation}
where we used the notation
$$H_{(i,3)}(y)=e^{-z_{i} y }-e^{- z_{3} y} \quad i=1,2$$
$$E_{i}(y)=e^{-z_{i} y}\quad i=1,2 ;\quad E_{3}(y)=e^{-z_{3} y}$$
The joining conditions for step n are 
\begin{equation}
  \begin{array}{c}
    A_{\pm}^{(n+1)} = E_{1}( y_{\mp}^{(n)})  A_{\mp}^{(n)} + w_1   \\
    B_{\pm}^{(n+1)} = E_{2}( y_{\mp}^{(n)})  B_{\mp}^{(n)} + w_2   
\end{array}
\label{eq:R}
\end{equation}
where \(w_1,  w_2\) are rational functions of the solutions \(z_1, z_2
z_3\). 

A simple (periodic) metachronal solution exists when the
transformation has a fixed point, which can be imposed setting the
equality between \(A_{+}^{(n+1)}\) and \(A_{+}^{(n)}\). This wave is
characterized by a unique \( y_+ = y_- = \bar{y} \) solution of  
\begin{displaymath}
  \frac{w_1}{1+E_{1}  (\bar{y})} H_{(1,3)}(\bar{y})  +
  \frac{w_{2}}{1+E_{2}(\bar{y})} H_{(2,3)}(\bar{y})  - 
  (1+\textrm{s}) E_{3} (\bar{y})+ (1-\textrm{s}) = 0
\end{displaymath}
and by coefficients  \( A_{-} = A_{+} \equiv \bar{A} , \quad
B_{-} = B_{+} \equiv \bar{B}   \) given by
$$ \bar{A} =\quad \frac{w_{1}}{1+ e^{-z_{1} \bar{y}}}  $$
$$ \bar{B} =\quad \frac{w_{2}}{1+ e^{-z_{2} \bar{y}}} $$
The equation for \(\bar{y}\) admits solution only if s \( < 1\), and,
for any given value of s, if \( p \) is lower than the critical value
\(p_c < \bar{p} \) introduced in the paper, which can be found
numerically.  This sets a maximal wavelength \(\lambda _{c}\) for the
metachronal wave.  The stability of the solution can be evaluated
linearizing the flow, starting from the point \( (\bar{A} + \delta A ,
\bar{B} + \delta B) \) in parameter space, inverting equation
\ref{eq:t+-} for \(\bar{y}\) and calculating the total variation
\((\delta A, \delta B)\) from \ref{eq:R}. In this case this yields one
negative and one positive eigenvalue corresponding to a marginally
stable fixed point.  The procedure outlined in this appendix can be
carried out more in general, leading to the results discussed in the
body of the paper.

\end{document}